\documentclass[conference]{IEEEtran}
\IEEEoverridecommandlockouts
\usepackage{algpseudocode}
\usepackage{algorithm}

\usepackage{cite}
\usepackage{amsmath,amssymb,amsfonts}
\usepackage{graphicx}
\usepackage{textcomp}
\usepackage{xcolor}
\def\BibTeX{{\rm B\kern-.05em{\sc i\kern-.025em b}\kern-.08em
    T\kern-.1667em\lower.7ex\hbox{E}\kern-.125emX}}
\begin{document}

\title{A QPINN Framework with Quantum Trainable Embeddings for the Lid-Driven Cavity Problem\\
\thanks{The authors acknowledge the support of the Danish e-Infrastructure Consortium (DeiC) and the National Quantum Algorithm Academy (NQAA) through the Postdoctoral Scholarship under the project ``Quantum-Driven Solutions for Multi-Agent Systems and Advanced Computation''. This work was also partially supported by UID/00147- Research Center for Systems and Technologies (SYSTEC) - and the Associate Laboratory Advanced Production and Intelligent Systems (ARISE, 10.54499/LA/P/0112/2020) funded by Fundação para a Ciência e a Tecnologia, I.P./ MCTES through the national funds. Furthermore, this research was supported, in part, by the Advanced Computing and Data Resource program, supported by the U.S. National Science Foundation (ACCESS), and Texas Tech University. The authors also acknowledge the research support provided by FPT University in Vietnam.}
}

\author{\IEEEauthorblockN{Nahid Binandeh Dehaghani}
\IEEEauthorblockA{\textit{Department of Electronic Systems} \\
\textit{Aalborg University}\\
Aalborg, Denmark \\
nahidbd@es.aau.dk}
\and
\IEEEauthorblockN{Ban Q. Tran}
\IEEEauthorblockA{\textit{Department of Computer Science} \\
\textit{Texas Tech University}\\
Lubbock, USA 
}
\IEEEauthorblockA{\textit{Department of Computing Fundamentals} \\
\textit{FPT University} \\
 Hanoi, Vietnam \\
 bantran@ttu.edu, bantq3@fe.edu.vn}
\and
\IEEEauthorblockN{Susan Mengel}
\IEEEauthorblockA{\textit{Department of Computer Science} \\
\textit{Texas Tech University}\\
Lubbock, USA \\
susan.mengel@ttu.edu}
\and
\IEEEauthorblockN{Rafal Wisniewski}
\IEEEauthorblockA{\textit{\quad \quad \quad Department of Electronic Systems} \\
\textit{\quad \quad Aalborg University}\\
\quad \quad Aalborg, Denmark \\
\quad \quad raf@es.aau.dk}
\and
\IEEEauthorblockN{A. Pedro Aguiar}
\IEEEauthorblockA{\textit{SYSTEC, ARISE, Faculty of Engineering} \\
\textit{University of Porto}\\
Porto, Portugal \\
pedro.aguiar@fe.up.pt}
}

\maketitle

\begin{abstract}
The steady incompressible Navier--Stokes equations pose significant
computational challenges due to their nonlinear convective terms and
pressure--velocity coupling. Physics-informed neural networks (PINNs)
provide a mesh-free framework for approximating such systems, but
classical PINNs can experience optimization difficulties in nonlinear
flow regimes. In this work, we propose a quantum physics-informed neural
network (QPINN) framework with a quantum neural network (QNN)-based
trainable embedding for the lid-driven cavity problem. The proposed
approach uses a QNN to learn data-adaptive quantum feature maps that
encode spatial coordinates before they are processed by a variational
quantum circuit within a physics-informed loss formulation. Numerical
experiments show that the proposed QNN-TE-QPINN exhibits stable training
behavior and competitive solution accuracy compared with classical PINNs
and hybrid quantum models using classical embeddings, while requiring
significantly fewer trainable parameters. Rather than claiming
computational speedup, these results highlight the potential of
trainable quantum embeddings for parameter-efficient physics-informed
learning. The findings suggest that embedding design plays an important
role in quantum-assisted PDE solvers and support further investigation
of QNN-based trainable embeddings for nonlinear fluid dynamics
benchmarks.
\end{abstract}


\section{Introduction}
\label{Introduction}

The numerical solution of nonlinear partial differential equations (PDEs)
remains one of the central challenges in scientific computing. Among
these equations, the incompressible Navier--Stokes equations play a
fundamental role in modeling fluid motion across science and
engineering, yet their nonlinear convective structure and strong
pressure--velocity coupling make them particularly difficult to solve
accurately. The lid-driven cavity flow is one of the most widely studied
benchmark problems in computational fluid dynamics (CFD), and it has
long served as a standard test case for assessing the robustness,
accuracy, and efficiency of numerical solvers for incompressible flow.
As demonstrated in the classical benchmark study of Ghia, Ghia, and
Shin~\cite{ghia1982high} and in subsequent investigations, the square
cavity geometry is simple, yet the resulting flow exhibits sufficiently
rich dynamics to pose a demanding challenge for numerical methods.

A primary source of difficulty in solving the lid-driven cavity problem
arises from the nonlinear convection terms in the Navier--Stokes
equations, which introduce self-interaction within the velocity field
and lead to complex flow features such as vortices and boundary-layer
formation~\cite{khorasanizade2014detailed}. Moreover, no closed-form
analytical solution exists for this problem, and accurate simulation
typically requires fine spatial discretization, particularly at higher
Reynolds numbers, where multiple vortical structures develop. These
computational demands motivate the exploration of alternative
approximation frameworks that can capture nonlinear flow behavior
without relying exclusively on conventional mesh-based techniques.

Physics-informed neural networks (PINNs), introduced by Raissi,
Perdikaris, and Karniadakis~\cite{raissi2019physics}, provide one such
framework by embedding governing equations, boundary conditions, and
initial conditions directly into the loss function of a neural network.
In this manner, PINNs learn continuous surrogate models of PDE solutions
while avoiding explicit discretization of the spatial domain.
Nevertheless, despite their conceptual elegance, classical PINNs often
encounter optimization challenges in strongly nonlinear and multiscale
settings, including slow convergence, sensitivity to loss balancing,
and difficulty in resolving complex solution manifolds.

These limitations have motivated interest in hybrid
quantum--classical approaches to physics-informed learning. In the noisy
intermediate-scale quantum (NISQ) era, variational quantum algorithms
based on parameterized quantum circuits have emerged as a practical
paradigm for near-term quantum computing~\cite{cerezo2021variational}.
Such approaches offer access to high-dimensional Hilbert-space
representations while remaining compatible with classical optimization
loops. Building on this idea, Kyriienko, Paine, and Elfving demonstrated
that differentiable quantum circuits can be trained to satisfy nonlinear
differential equations and boundary conditions, thereby opening a path
toward quantum-assisted PDE solvers~\cite{kyriienko2021solving}.

A key architectural component in quantum-assisted PINNs is the
embedding mechanism that maps classical input coordinates into quantum
states prior to variational processing. In the original
trainable-embedding QPINN framework, Berger, Hosters, and
M{\"o}ller~\cite{berger2025trainable} introduced adaptive, trainable
feature maps to overcome the expressivity limitations of fixed
encodings, showing that learned embeddings can improve solution accuracy
while reducing qubit requirements. Motivated by this insight, our recent
work has systematically investigated trainable embeddings in
quantum-assisted physics-informed learning, examining both classical and
fully quantum embedding strategies and their impact on convergence,
gradient structure, parameter scaling, and solution quality in nonlinear
reaction--diffusion systems~\cite{tran2026trainable,tran2026quantum,dehaghani2025quantum}.

In the present study, we extend this line of research to nonlinear fluid
dynamics by proposing a QPINN framework with a quantum neural network
(QNN)-based trainable embedding for the lid-driven cavity problem. In
this architecture, a dedicated QNN learns a mapping from collocation
point coordinates to quantum data-encoding angles, which are then
processed by a variational quantum circuit to approximate the solution
fields. 
Unlike fixed or handcrafted encodings, the proposed approach learns the coordinate-to-angle mapping jointly with the PDE solver. This gives the quantum circuit an adaptive input representation that can better match the spatial structure of the lid-driven cavity flow.

\paragraph*{Contributions and Scope}
The main contributions of this work are as follows. First, we formulate
a trainable-embedding quantum physics-informed neural network (QPINN)
framework for the steady incompressible Navier--Stokes equations
governing the lid-driven cavity problem. Second, we introduce a
QNN-based trainable embedding mechanism that learns a coordinate-to-angle encoding transformation jointly with the variational quantum solver.
Third, we provide an empirical comparison with classical PINNs and
hybrid QPINN baselines employing alternative embedding strategies.
Finally, through numerical experiments, we show that embedding design is
an important factor in quantum-assisted PDE solvers, extending the use
of trainable-embedding QPINNs to a nonlinear fluid-flow benchmark.


We do not claim a complexity-theoretic quantum advantage for the proposed method. Instead, the observed benefit is a reduction in the number of trainable parameters required to achieve comparable accuracy. In this sense, parameter efficiency means that the model has fewer degrees of freedom to optimize, which can reduce memory requirements, lower optimizer overhead, and make the training problem more compact. In the reported experiments, the proposed QNN-TE-QPINN achieves competitive accuracy for the lid-driven cavity problem using approximately $360$ trainable parameters, compared with approximately $6{,}600$ parameters in the classical PINN baseline. This comparison should therefore be understood as an empirical reduction in model size and trainable degrees of freedom, rather than as a demonstrated computational speedup.

\section{Navier--Stokes Equations in Stream-Function Formulation}

In this work, we consider the steady two-dimensional incompressible Navier--Stokes equations for the lid-driven cavity problem on the square domain $\Omega=[0,1]\times[0,1]$. Let $u(x,y)$ and $v(x,y)$ denote the horizontal and vertical velocity components, respectively, and let $p(x,y)$ denote the pressure field. In the nondimensional steady-state setting considered here, the momentum equations are given by
\begin{equation}
u\frac{\partial u}{\partial x}+v\frac{\partial u}{\partial y}
=
-\frac{\partial p}{\partial x}
+\frac{1}{Re}\left(\frac{\partial^2 u}{\partial x^2}+\frac{\partial^2 u}{\partial y^2}\right),
\label{eq:ns_momentum_x}
\end{equation}
\begin{equation}
u\frac{\partial v}{\partial x}+v\frac{\partial v}{\partial y}
=
-\frac{\partial p}{\partial y}
+\frac{1}{Re}\left(\frac{\partial^2 v}{\partial x^2}+\frac{\partial^2 v}{\partial y^2}\right),
\label{eq:ns_momentum_y}
\end{equation}
where $Re$ denotes the Reynolds number.

A major difficulty in solving the Navier--Stokes equations arises from
the nonlinear convective terms $u u_x + v u_y$ and $u v_x + v v_y$,
where subscripts denote partial derivatives, e.g.,
$u_x=\partial u/\partial x$ and $v_y=\partial v/\partial y$.
These nonlinear self-advection terms couple the unknown velocity field
to its own spatial derivatives. This nonlinear self-advection is
responsible for the rich behavior of incompressible flows, including
vortex formation and increased sensitivity to boundary conditions as
the Reynolds number increases.
For two-dimensional incompressible flows on a simply connected domain, the divergence-free condition implies the existence of a scalar stream function $\psi(x, y)$. Therefore, we adopt the stream-function formulation

\begin{equation}
u=\frac{\partial \psi}{\partial y},
\qquad
v=-\frac{\partial \psi}{\partial x}.
\label{eq:stream_function_def}
\end{equation}
With this definition, the divergence-free condition is satisfied identically:
\begin{equation}
\frac{\partial u}{\partial x}+\frac{\partial v}{\partial y}
=
\frac{\partial^2 \psi}{\partial x\partial y}
-\frac{\partial^2 \psi}{\partial y\partial x}=0.
\label{eq:stream_div_free}
\end{equation}
Therefore, instead of learning the velocity components directly, the model approximates the pair $(p,\psi)$, and the velocity field is recovered through~\eqref{eq:stream_function_def}. This is the formulation used in the implementation, where incompressibility is enforced implicitly through the stream function rather than through a separate continuity residual in the loss function.

Using~\eqref{eq:stream_function_def}, the momentum equations are enforced in residual form as
\begin{equation}
\mathcal{R}_x :=
u\frac{\partial u}{\partial x}
+
v\frac{\partial u}{\partial y}
+
\frac{\partial p}{\partial x}
-
\frac{1}{Re}\left(
\frac{\partial^2 u}{\partial x^2}
+
\frac{\partial^2 u}{\partial y^2}
\right)=0,
\label{eq:residual_x}
\end{equation}
\begin{equation}
\mathcal{R}_y :=
u\frac{\partial v}{\partial x}
+
v\frac{\partial v}{\partial y}
+
\frac{\partial p}{\partial y}
-
\frac{1}{Re}\left(
\frac{\partial^2 v}{\partial x^2}
+
\frac{\partial^2 v}{\partial y^2}
\right)=0,
\label{eq:residual_y}
\end{equation}
where $u$ and $v$ are obtained from derivatives of $\psi$.

For the lid-driven cavity problem, the boundary conditions are defined as follows. On the top boundary, the lid moves with unit horizontal velocity and zero vertical velocity:
\begin{equation}
u(x,1)=1,
\qquad
v(x,1)=0.
\label{eq:lid_bc}
\end{equation}
On the remaining three walls, define
$\partial\Omega_{\mathrm{wall}}
:= \{(x,y)\in\Omega \mid x=0 \ \text{or}\ x=1 \ \text{or}\ y=0\}$.
The no-slip boundary conditions are imposed as
\begin{equation}
u=0,
\qquad
v=0,
\quad \text{on } \partial\Omega_{\mathrm{wall}}.
\label{eq:noslip_bc}
\end{equation}
Since pressure is determined only up to an additive constant in incompressible flow, a reference pressure is prescribed to ensure uniqueness. In this work, we impose
\begin{equation}
p(0,0)=0.
\label{eq:pressure_ref}
\end{equation}
Hence, the learning problem consists of approximating $p(x,y)$ and $\psi(x,y)$ such that the induced velocity field from~\eqref{eq:stream_function_def} satisfies the momentum residuals~\eqref{eq:residual_x}--\eqref{eq:residual_y}, together with the lid-driven and no-slip boundary conditions~\eqref{eq:lid_bc}--\eqref{eq:noslip_bc} and the reference pressure condition~\eqref{eq:pressure_ref}. This formulation matches the computational structure used in our solver.

\begin{figure}[t]
\includegraphics[scale=0.125]{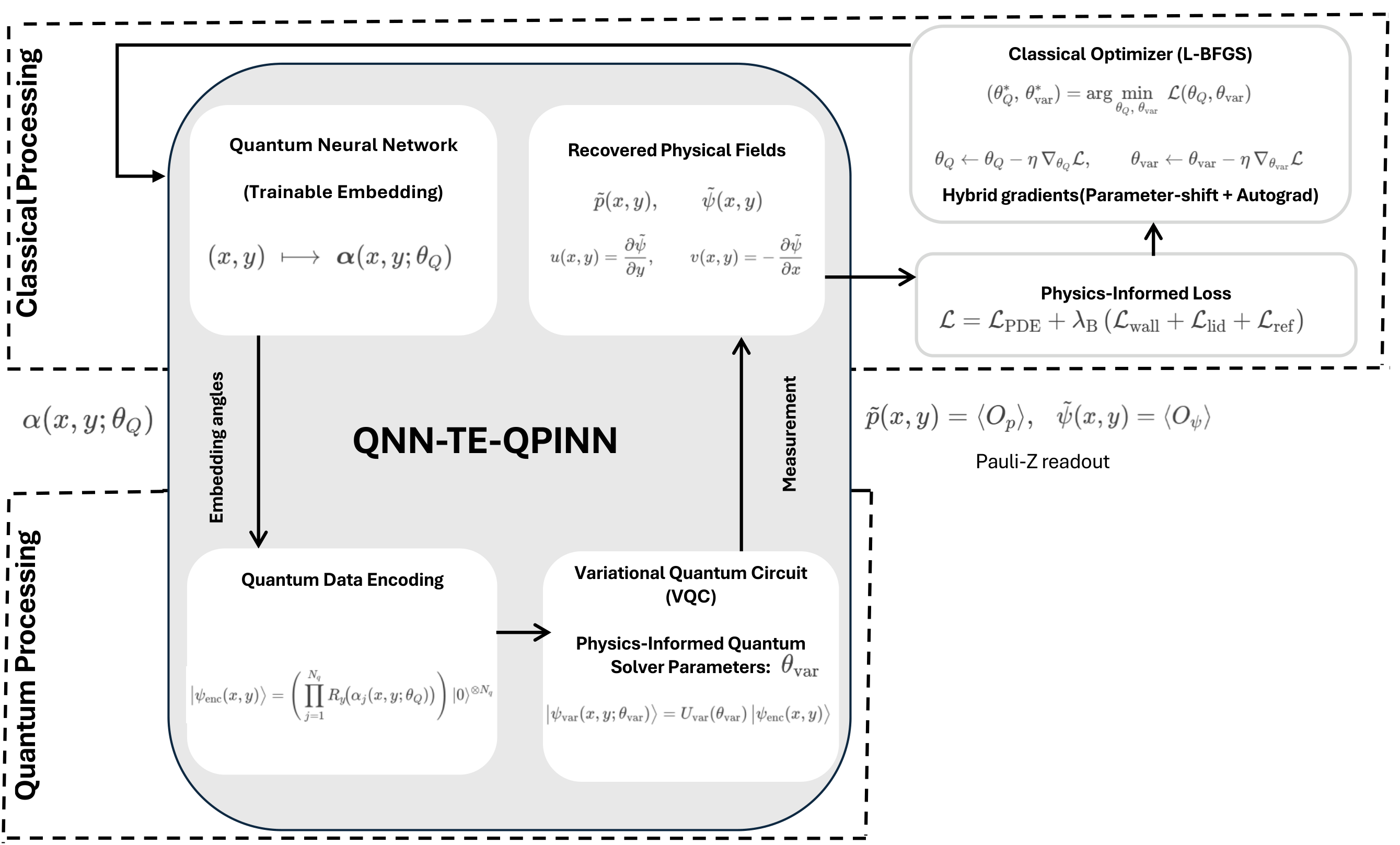}
\caption{Architecture of the proposed QNN-TE-QPINN framework. Classical
processing maps spatial collocation points $(x,y)$ to trainable embedding
parameters $\boldsymbol{\alpha}(x,y;\theta_Q)$ via a quantum neural
network. These parameters are used for quantum data encoding, after
which a variational quantum circuit acts as the physics-informed
quantum solver to generate a variational quantum state. Observable
expectation values obtained through Pauli--$Z$ measurements yield
predictions for the pressure $p(x,y)$ and stream function $\psi(x,y)$,
from which velocity fields are recovered. Physics-informed PDE
residuals, boundary constraints, and reference conditions are enforced
through a classical loss function. All parameters are updated using a
classical L-BFGS optimizer with hybrid gradients computed via automatic
differentiation and parameter-shift rules.}
\label{qnn_arch}
\end{figure}

\section{Quantum Physics-Informed Learning Framework}
In this section, we introduce the hybrid quantum--classical physics-informed learning framework used to approximate the steady incompressible Navier--Stokes equations in the lid-driven cavity problem. The proposed approach extends the trainable-embedding quantum PINN paradigm to nonlinear fluid dynamics by combining (i) quantum neural network (QNN)-based embeddings, (ii) a hardware-efficient variational quantum circuit, and (iii) a physics-informed loss that enforces momentum balance, boundary conditions, and pressure normalization. Figure~\ref{qnn_arch} provides an overview of the computational pipeline.

\subsection{Trainable Quantum Embedding Network (QNN-Based)}

To encode the spatial coordinates $(x,y)$ of the cavity domain into a
form suitable for quantum processing, we employ a trainable quantum
embedding network. This module translates classical inputs into a set of
quantum features that determine how the data enter the variational
quantum circuit, and it provides the only channel through which the
physical coordinates influence the quantum model. As a result, the
embedding circuit defines a data-dependent quantum feature map whose
structure is learned jointly with the PDE solution.

Before being embedded, the coordinates are rescaled through an affine
mapping,
\begin{equation}
    \tilde{x} = \mathcal{N}(x), \qquad 
    \tilde{y} = \mathcal{N}(y),
\end{equation}
which maps the original coordinate range to the normalized interval
$[-1,1]$. This normalization balances the magnitudes of the two spatial
dimensions and improves the numerical stability of the hybrid
optimization process.
The normalized coordinates $(\tilde{x},\tilde{y})$ are then processed by
a parameterized quantum circuit equipped with variational parameters
$\theta_Q$. Acting on the reference state $|0\rangle^{\otimes N_q}$,
this circuit produces the embedded quantum state
\begin{equation}
    |\psi_{\mathrm{embed}}(\tilde{x},\tilde{y};\theta_Q)\rangle
    =
    U_{\mathrm{embed}}(\tilde{x},\tilde{y};\theta_Q)
    |0\rangle^{\otimes N_q}.
\end{equation}
The unitary $U_{\mathrm{embed}}$ consists of alternating layers of 
input-dependent gates---through which the coordinates affect the
state---and trainable rotation layers parameterized by $\theta_Q$, along
with a predefined pattern of entangling gates, such as nearest-neighbor CNOT gates.
This design enables the embedding
map to adapt its representation during training and to distribute
coordinate information across the qubits in a highly nonlinear manner.

To convert the embedded quantum state into a usable feature vector, we
evaluate a prescribed set of Hermitian observables
$\{O_i\}_{i=1}^{N_q}$, obtaining
\begin{equation}
    \Gamma(\tilde{x},\tilde{y})
    =
    \big[
        \alpha_1(\tilde{x},\tilde{y}),\,
        \alpha_2(\tilde{x},\tilde{y}),\,
        \ldots,\,
        \alpha_{N_q}(\tilde{x},\tilde{y})
    \big]^{\!\top},
\end{equation}
where
   $ \alpha_i(\tilde{x},\tilde{y})
    =
    g_i\!\left(
        \langle 
            \psi_{\mathrm{embed}}(\tilde{x},\tilde{y};\theta_Q)
        |
            O_i
        |
            \psi_{\mathrm{embed}}(\tilde{x},\tilde{y};\theta_Q)
        \rangle
    \right),
$
and $g_i(\cdot)$ is a scaling function that maps the measurement outcome
to a valid rotation angle. In this work, we take $O_i$ to be the Pauli--$Z$ operator on the $i$-th
qubit, and the post-processing function reduces to a simple linear
scaling $g_i(s)=\pi s$, which maps the expectation values in $[-1,1]$ to
rotation angles used in the quantum encoding unitary.
This choice provides a simple yet expressive quantum feature map, since
Pauli--$Z$ measurements yield bounded outputs in $[-1,1]$, and the linear
scaling ensures that the embedding angles span the full rotation range
required for downstream encoding.
Therefore, the QNN embedding learns a nonlinear coordinate transformation
tailored to the flow physics, enabling the quantum model to represent
vorticity, shear, and pressure--velocity coupling using a compact and
hardware-efficient architecture.

\paragraph*{Quantum encoding operator}
The embedding circuit produces classicalized quantum features, which are
then re-encoded into the solver circuit through a standard rotational
data-encoding unitary.
The angles $\alpha_i$ determine the data-encoding unitary applied prior
to the variational circuit,
\begin{equation}
    U_{\mathrm{enc}}(\tilde{x},\tilde{y};\theta_Q)
    =
    \bigotimes_{i=1}^{N_q}
    R_y\!\left(\alpha_i(\tilde{x},\tilde{y})\right),
\end{equation}
where $R_y(\cdot) = \exp[-\,i(\cdot)\sigma_y/2]$ denotes a single-qubit
rotation about the $y$-axis. This operator produces the encoded
input-dependent quantum state
\begin{equation}
    |\psi_{\mathrm{enc}}(\tilde{x},\tilde{y})\rangle
    =
    U_{\mathrm{enc}}(\tilde{x},\tilde{y};\theta_Q)
    |0\rangle^{\otimes N_q}.
\end{equation}
Because the embedding parameters $\theta_Q$ are optimized jointly with
the variational parameters of the downstream circuit, the feature map
$\Gamma(\tilde{x},\tilde{y})$ adapts dynamically to the structure of the
Navier--Stokes residuals. This co-training strategy enables the embedding
to learn representations that emphasize key flow characteristics---such as
vorticity concentration, shear layers, and coupled pressure--velocity
behavior---while the use of shallow, hardware-efficient quantum circuits
ensures compatibility with near-term devices. The result is a flexible
and fully quantum trainable embedding mechanism that provides a
consistent representation of the domain and can enhance the
expressiveness of the overall QPINN model.

\subsection{Variational Quantum Circuit Design}

After the embedding module produces an input-dependent quantum state,
this state is passed through a parameterized variational quantum circuit
(VQC), which forms the trainable quantum component of the hybrid learning
architecture. The purpose of the VQC is to transform the encoded state
into a latent quantum representation from which the physical output
fields can be obtained through measurement.

The variational circuit is governed by a set of trainable parameters
$\theta_{\mathrm{var}}$ and is applied uniformly to all spatial inputs.
That is, the same variational unitary acts repeatedly on quantum states
generated from different coordinate-dependent embeddings. This
parameter-sharing strategy lowers the number of trainable quantum
variables and enforces a single latent representation for the predicted
Navier--Stokes fields. Distinct physical quantities may then be extracted
from the resulting quantum state by measuring different observables.

The variational unitary is constructed as a sequence of $L$ layers,
$
    U_{\mathrm{var}}(\theta_{\mathrm{var}})
    =
    U_L(\theta_L)\cdots U_2(\theta_2)\,U_1(\theta_1),
$
and each layer $U_\ell(\theta_\ell)$ consists of a product of
parameterized and fixed operations. More precisely, each trainable block
is given by
\begin{equation}
    U_\ell(\theta_\ell)
    =
    \prod_m \exp\!\bigl(-\,\mathrm{i}\,\theta_{\ell,m} H_m\bigr)\, W_m ,
\end{equation}
where the generators $H_m$ are predetermined Hermitian operators and the
operators $W_m$ denote fixed, non-parameterized unitaries.

Each layer thus contains learnable single-qubit rotations, followed by a
fixed entangling pattern. In typical hardware-efficient designs, the
generators $H_m$ are chosen from the Pauli basis and correspond to
rotations about the coordinate axes of the Bloch sphere, while the
operators $W_m$ implement a nearest-neighbor entangling structure such
as a sequence of CNOT gates. This layered composition allows the
expressive power of the circuit to grow with depth while preserving a
shallow architecture suitable for near-term quantum devices.

For a given spatial location $(x,y)$, the variational quantum state is
obtained by applying the shared VQC to the encoded quantum state
generated by the data-encoding operator:
\begin{equation}
    |\psi_{\mathrm{var}}(x,y;\Theta)\rangle
    =
    U_{\mathrm{var}}(\theta_{\mathrm{var}})
    U_{\mathrm{enc}}(\tilde{x},\tilde{y};\theta_Q)
    |0\rangle^{\otimes N_q},
\end{equation}
where $\Theta=(\theta_Q,\theta_{\mathrm{var}})$ collects all trainable
parameters of the model.

The predicted pressure and stream-function fields are then obtained as
expectation values of the corresponding observables evaluated on the
state $|\psi_{\mathrm{var}}\rangle$. The variational parameters are
optimized simultaneously with the embedding parameters using the
physics-informed loss introduced in the following subsection. Repeated
application of the same VQC to quantum states encoding different points
in the domain enables the model to capture the nonlinear interactions,
spatial gradients, and flow structures characteristic of incompressible
Navier--Stokes dynamics, while maintaining a hardware-efficient
quantum circuit compatible with near-term implementations.

\subsection{Quantum Readout and Observable Design}
The hybrid quantum model produces predictions for the pressure and
stream-function fields by evaluating expectation values of suitably
chosen Hermitian observables acting on the variational quantum state.
While the embedding and variational circuits generate a latent quantum
representation of the spatial coordinates, distinct physical fields are
extracted at the readout stage through field-specific observable
operators.

For each output field $i \in \{p,\psi\}$, we assign a corresponding
readout operator $O_i$ acting on the $N_q$-qubit Hilbert space. In this
work, we employ Pauli--$Z$-based observables of the form
$
    O_i = \sum_{j=1}^{N_q} Z_j ,
$
where $Z_j$ denotes the Pauli--$Z$ operator acting on the $j$-th qubit.
This choice yields a single scalar expectation value per circuit
evaluation, providing an efficient and hardware-friendly mechanism for
mapping quantum states to classical outputs.
Pauli--$Z$ observables are particularly advantageous in
physics-informed learning, as their bounded spectrum leads to stable
expectation values. This boundedness improves the numerical conditioning
of the PDE residuals and their derivatives, which in turn mitigates
training instabilities and oscillatory behavior during optimization.
The predicted pressure and stream-function fields are therefore computed
as
\begin{equation}
    \tilde{p}(x,y)
    =
    \big\langle
        \psi_{\mathrm{var}}(x,y;\Theta)
        \,\big|\,
        O_p
        \,\big|\,
        \psi_{\mathrm{var}}(x,y;\Theta)
    \big\rangle,
\end{equation}
\begin{equation}
    \tilde{\psi}(x,y)
    =
    \big\langle
        \psi_{\mathrm{var}}(x,y;\Theta)
        \,\big|\,
        O_\psi
        \,\big|\,
        \psi_{\mathrm{var}}(x,y;\Theta)
    \big\rangle,
\end{equation}
where $|\psi_{\mathrm{var}}(x,y;\Theta)\rangle$ denotes the variational
quantum state produced by the embedding and variational circuits, and
$O_p$ and $O_\psi$ are readout operators associated with the pressure and
stream-function fields, respectively. The specific solver architecture
adopted in the experiments---including whether readouts are obtained from
a shared variational circuit or from independent solver instances---is
detailed in the simulation section.
While Pauli--$Z$ readouts provide a robust and low-overhead extraction
strategy, the proposed framework readily supports more general Hermitian
observables, such as weighted Pauli combinations or problem-dependent
operators, provided they remain efficiently measurable on near-term
quantum hardware. The choice of readout operator thus reflects a balance
between expressiveness, numerical stability, and practical feasibility
in quantum-assisted solutions of the incompressible Navier--Stokes
equations.

\subsection{Physics-Informed Loss Function}

The trainable parameters of the QNN-TE-QPINN model are optimized by
minimizing a physics-informed loss function that enforces the steady
incompressible Navier--Stokes equations together with the associated
boundary and reference constraints. The loss is composed of four terms:
the interior momentum residuals, the no-slip wall conditions, the
lid-driven boundary condition, and a reference pressure constraint.

\paragraph*{PDE residual loss}
At interior collocation points, the model predicts the pressure and
stream-function fields $(p,\psi)$, from which the velocity components
are recovered via $u=\partial\psi/\partial y$ and
$v=-\partial\psi/\partial x$. Substituting these expressions into the
momentum equations yields the residuals
\begin{equation}
    \mathcal{R}_x(x,y)
    =
    u\,u_x + v\,u_y
    + p_x
    - \frac{1}{Re}\left(u_{xx} + u_{yy}\right),
\end{equation}
\begin{equation}
    \mathcal{R}_y(x,y)
    =
    u\,v_x + v\,v_y
    + p_y
    - \frac{1}{Re}\left(v_{xx} + v_{yy}\right),
\end{equation}
where all spatial derivatives are computed using automatic differentiation through the hybrid quantum–classical model. The interior physics loss is defined as the mean squared magnitude
of the momentum residuals,
\begin{equation}
    \mathcal{L}_{\mathrm{PDE}}
    =
    \frac{1}{|\Omega_{\mathrm{int}}|}
    \sum_{(x,y)\in\Omega_{\mathrm{int}}}
    \left(
        \mathcal{R}_x(x,y)^2
        +
        \mathcal{R}_y(x,y)^2
    \right).
\end{equation}

\paragraph*{Boundary losses}
The no-slip boundary condition requires $u=0$ and $v=0$ on the stationary
walls, while the moving lid enforces $u=1$ and $v=0$ at $y=1$. These
constraints are imposed weakly by penalizing the squared deviations of
the predicted velocities at the corresponding boundary collocation
points,
\begin{equation}
    \mathcal{L}_{\mathrm{wall}}
    =
    \frac{1}{|\partial\Omega_{\mathrm{wall}}|}
    \sum_{(x,y)\in\partial\Omega_{\mathrm{wall}}}
    \left(
        u(x,y)^2 + v(x,y)^2
    \right),
\end{equation}
\begin{equation}
    \mathcal{L}_{\mathrm{lid}}
    =
    \frac{1}{|\partial\Omega_{\mathrm{lid}}|}
    \sum_{(x,y)\in\partial\Omega_{\mathrm{lid}}}
    \left(
        (u(x,y)-1)^2 + v(x,y)^2
    \right).
\end{equation}

\paragraph*{Reference pressure constraint}
Since the pressure field in incompressible flow is defined only up to an
additive constant, a reference value is prescribed to ensure uniqueness.
In this work, we enforce the normalization condition $p(0,0)=0$ through
the penalty term
\begin{equation}
    \mathcal{L}_{\mathrm{ref}}
    =
    p(0,0)^2.
\end{equation}

\paragraph*{Total loss}
The overall physics-informed loss combines the interior residual with
the boundary and reference terms. To balance the relative contributions
of the different constraints during optimization, the boundary and
reference losses are scaled by a weighting parameter
$\lambda_{\mathrm{B}}$. The total loss is given by
   $ \mathcal{L}
    =
    \mathcal{L}_{\mathrm{PDE}}
    +
    \lambda_{\mathrm{B}}
    \left(
        \mathcal{L}_{\mathrm{wall}}
        +
        \mathcal{L}_{\mathrm{lid}}
        +
        \mathcal{L}_{\mathrm{ref}}
    \right).$
By jointly minimizing $\mathcal{L}$ with respect to all trainable
parameters $\Theta$, the hybrid quantum model learns pressure and
stream-function fields that satisfy the Navier--Stokes equations while
remaining consistent with the lid-driven cavity boundary conditions.

\subsection{Gradient Structure and Derivative Computation}

Training the QNN-TE-QPINN model requires computing derivatives of the
predicted pressure and stream-function fields with respect to both the
spatial coordinates and the trainable parameters of the hybrid
architecture. These derivatives enter directly into the evaluation of
the momentum residuals, which involve first- and second-order spatial
derivatives of the velocity components. As a result, the learning
procedure depends on an interplay between quantum differentiation of the
embedding and variational circuits and classical automatic
differentiation of the PDE operators.

Since the predicted fields $\tilde{p}(x,y)$ and $\tilde{\psi}(x,y)$ are
obtained as expectation values of the variational quantum state, their
dependence on $(x,y)$ arises exclusively through the data-encoding
angles generated by the quantum embedding network. By the chain rule,
spatial derivatives decompose into
\begin{equation}
    \frac{\partial \tilde{f}}{\partial x}
    =
    \sum_{m=1}^{N_q}
    \frac{\partial \tilde{f}}{\partial \alpha_m}
    \frac{\partial \alpha_m}{\partial x},
    \qquad
    f\in\{p,\psi\},
\end{equation}
where $\partial \tilde{f}/\partial \alpha_m$ denotes the sensitivity of
the quantum expectation value with respect to the rotation angle on the
$m$-th qubit, and $\partial \alpha_m/\partial x$ is obtained by
classical backpropagation through the QNN embedding network. An
analogous expression holds for $\partial \tilde{f}/\partial y$.
Second-order derivatives required by the Navier--Stokes operators
($u_{xx}, u_{yy}, v_{xx}, v_{yy}$) follow from repeated application of
the same chain rule and are computed through nested automatic
differentiation.

The gradients of the total loss with respect to any variational or
embedding parameter $\theta\in\Theta$ take the form
\begin{equation}
    \frac{\partial \mathcal{L}}{\partial \theta}
    =
    \sum_{(x,y)\in\Omega_{\mathrm{int}}}
    \left(
        \frac{\partial \mathcal{L}}{\partial \tilde{p}}
        \frac{\partial \tilde{p}}{\partial \theta}
        +
        \frac{\partial \mathcal{L}}{\partial \tilde{\psi}}
        \frac{\partial \tilde{\psi}}{\partial \theta}
    \right)
    + 
    \frac{\partial \mathcal{L}_{\mathrm{B}}}{\partial \theta},
\end{equation}
where $\mathcal{L}_{\mathrm{B}}$ denotes the collection of boundary and
reference terms. The derivatives
$\partial \tilde{p}/\partial \theta$ and
$\partial \tilde{\psi}/\partial \theta$ reduce to gradients of
quantum expectation values with respect to circuit parameters and are
evaluated through the differentiable quantum-classical interface. In the
present implementation, these gradients are computed using the automatic
differentiation capabilities of the underlying hybrid framework, which
internally applies parameter-shift rules for quantum gates and classical
backpropagation for the embedding network.

Because the momentum residuals admit a closed-form dependence on the
predicted fields and their spatial derivatives, the derivatives
$\partial \mathcal{L}/\partial \tilde{p}$ and
$\partial \mathcal{L}/\partial \tilde{\psi}$ are obtained directly via
automatic differentiation. The resulting optimization pipeline therefore
combines classical autograd for the PDE components with quantum-aware
gradient evaluation for the circuit parameters, enabling the hybrid
model to capture the nonlinear coupling, steep gradients, and boundary
layer behavior characteristic of the lid-driven cavity flow.

\begin{algorithm}
\small
\caption{Training Algorithm for the QNN-TE-QPINN Framework (Navier--Stokes)}
\label{alg:qnnteqpinn}

\begin{algorithmic}[1]

\State \textbf{Input:}
\Statex \quad Navier--Stokes momentum operators $\mathcal{R}_x[\cdot], \mathcal{R}_y[\cdot]$
\Statex \quad Boundary conditions: no-slip walls and lid-driven boundary
\Statex \quad Pressure reference constraint $p(0,0)=0$
\Statex \quad Collocation sets $\mathcal{S}_{\mathrm{int}}, \mathcal{S}_{\mathrm{wall}}, \mathcal{S}_{\mathrm{lid}}$
\Statex \quad Boundary weight $\lambda_{\mathrm{B}}$
\Statex \quad Maximum iterations $N_{\max}$, tolerance $\varepsilon$
\Statex \quad Embedding type: QNN-based

\State \textbf{Initialization:}
\State Initialize quantum embedding parameters $\theta_{Q}$
\State Initialize variational circuit parameters $\theta_{\mathrm{var}}$
\State Select readout observables $O_p, O_{\psi}$
\State Normalize spatial coordinates $(x,y)$ to $(\tilde{x},\tilde{y}) \in [-1,1]^2$
\State Initialize optimizer for $(\theta_{Q},\theta_{\mathrm{var}})$

\While{not converged}

\State \textbf{Forward evaluation}

\For{each collocation point $(x^j,y^j)$}

    \State Compute embedding angles 
    $\alpha(x^j,y^j;\theta_Q)$

    \State Prepare encoded quantum state
$
    |\psi_{\mathrm{enc}}\rangle
    = U_{\mathrm{enc}}(\alpha)\,|0\rangle^{\otimes N_q}
$
    \State Apply variational circuit
    $
        |\psi_{\mathrm{var}}\rangle
        =
        U_{\mathrm{var}}(\theta_{\mathrm{var}})
        |\psi_{\mathrm{enc}}\rangle
    $

    \State Evaluate predicted fields
    $
        \tilde{p}(x^j,y^j)=
        \langle \psi_{\mathrm{var}}|O_p|\psi_{\mathrm{var}}\rangle,$
     $   \tilde{\psi}(x^j,y^j)=
        \langle \psi_{\mathrm{var}}|O_{\psi}|\psi_{\mathrm{var}}\rangle
    $

    \State Compute velocity components
    $
        u=\frac{\partial\tilde{\psi}}{\partial y},$
     $   v=-\frac{\partial\tilde{\psi}}{\partial x}
    $

    \State Compute all spatial derivatives of 
    $(u,v,p)$ via hybrid differentiation

\EndFor

\State \textbf{Loss evaluation}

\State Compute PDE residual loss $\mathcal{L}_{\mathrm{PDE}}$

\State Compute no-slip loss 
$\mathcal{L}_{\mathrm{wall}}$

\State Compute lid-driven boundary loss 
$\mathcal{L}_{\mathrm{lid}}$

\State Compute pressure reference loss 
$\mathcal{L}_{\mathrm{ref}}=p(0,0)^2$

\State Total loss:
$
\mathcal{L}
=
\mathcal{L}_{\mathrm{PDE}}
+
\lambda_{\mathrm{B}}
\left(
    \mathcal{L}_{\mathrm{wall}}
    +
    \mathcal{L}_{\mathrm{lid}}
    +
    \mathcal{L}_{\mathrm{ref}}
\right)
$

\State \textbf{Gradient computation}

\State Compute $\nabla_{\theta_{Q}} \mathcal{L}$ via hybrid differentiation
(classical backprop through the embedding network and quantum
parameter-shift for embedding gates)

\State Compute $\nabla_{\theta_{\mathrm{var}}} \mathcal{L}$ via quantum
parameter‑shift rule

\State \textbf{Parameter update}

\State Update $(\theta_{Q},\theta_{\mathrm{var}})$ with selected optimizer

\EndWhile

\State \textbf{Output:}
\State Trained parameters $(\theta_{Q}^{\star},\theta_{\mathrm{var}}^{\star})$ and learned fields $\tilde{p}(x,y)$ and $\tilde{\psi}(x,y)$

\end{algorithmic}
\end{algorithm}

\subsection{Computational Cost and Parameter Scaling}
The training cost of the QNN-TE-QPINN architecture arises from a
combination of classical and quantum components, each contributing
distinctly to the overall computational load. For each training
iteration, the dominant expense comes from evaluating the hybrid model at
all collocation points and computing the required gradients through both
the embedding network and the variational circuit.

On the quantum side, the cost scales with the number of qubits
$N_q$ and the depth $L$ of the variational circuit. Each trainable
rotation gate introduces a parameter whose gradient must be computed
through quantum-aware differentiation, and thus the total number of
quantum parameters typically grows linearly with $N_qL$. Because the training
procedure requires evaluating the quantum circuit once per collocation
point, the total quantum evaluation cost scales approximately as
$N_{\mathrm{col}}N_qL$, where $N_{\mathrm{col}}$ denotes the number of
collocation points. When using a QNN-based embedding, additional quantum
evaluations are required to compute the embedding angles, contributing a
cost proportional to the depth of the embedding circuit. In contrast,
classical embeddings incur no quantum overhead, but require additional
backpropagation through a classical network.

The classical computational cost includes evaluating the PDE residuals,
computing all first- and second-order spatial derivatives via automatic
differentiation, enforcing the boundary and reference constraints, and
computing the loss gradients. These classical operations scale with
$N_{\mathrm{col}}$ and with the complexity of the derivative chain-rule
structure in the Navier--Stokes system. Although the Navier--Stokes
residuals are more expensive to evaluate than those of scalar PDEs, the
cost remains polynomial in the number of collocation points.

The parameter count of the model grows polynomially with
$N_q$, $L$, and the depth of the embedding circuit. This ensures that
the hybrid quantum architecture remains trainable with standard gradient
optimizers, while allowing the expressive capacity of the model to be
tuned according to the available quantum resources. The separation
between embedding and variational parameters further provides flexibility
in distributing computational effort between classical preprocessing and
quantum state preparation, making the architecture compatible with
near-term quantum devices.

\subsection{Representational Capacity and Parameter Efficiency}

The QNN-TE-QPINN framework leverages the expressive power of
parameterized quantum circuits to approximate the nonlinear solution
manifold of the Navier--Stokes equations. Each variational quantum
circuit acts on an $N_q$-qubit Hilbert space of dimension $2^{N_q}$,
providing a high-dimensional latent representation in which nonlinear
transformations can be implemented using relatively shallow circuits.
Although the model does not attempt to explore the full Hilbert space,
the exponentially large state space provides structural flexibility that
may be advantageous for encoding complex flow features.

The combination of a trainable quantum embedding and a shared
variational circuit allows the model to adaptively map the spatial
coordinates into quantum states that highlight relevant flow structures,
including boundary layers, shear regions, and vortex formation.
Entanglement generated by the variational layers introduces correlations
across qubits, enabling the quantum model to represent multi-scale
interactions and nonlinear coupling between pressure and velocity.
Because the Navier--Stokes system is characterized by strongly
nonlinear advection and spatial stiffness near solid boundaries, the
ability of quantum circuits to implement global transformations in a
compact form may offer a representational advantage over comparable
classical networks.

At the same time, the architecture remains compatible with near-term
quantum devices through the use of hardware-efficient ansätze and shallow
embedding circuits. The model balances expressiveness and resource cost
by restricting parameter growth to linear scaling in $N_q$ and $L$,
rather than relying on deep or highly entangled circuits. As quantum
hardware improves, this structural expressivity may support more
accurate and efficient quantum-assisted solvers for fluid dynamics and
other nonlinear PDEs, while preserving a modular hybrid design that
integrates smoothly with classical automatic differentiation and
physics-informed optimization.

\section{Numerical Results}
\label{Ex-Results}
In this section, we evaluate the proposed QNN-TE-QPINN
framework for the steady incompressible Navier--Stokes
equations associated with the lid-driven cavity problem. The
experimental study is designed to assess the training behavior,
solution quality, and generalization capability of the proposed
quantum-assisted model. We first describe the experimental
setup, including the computing environment, model
configurations, quantum circuit architectures, collocation
strategy, and reference solutions. We then compare different
embedding strategies, analyze the training behavior against
classical and hybrid quantum baselines, and report inference
results using quantitative error metrics.

\begin{figure}[htbp]
\centerline{\includegraphics[scale=0.18]{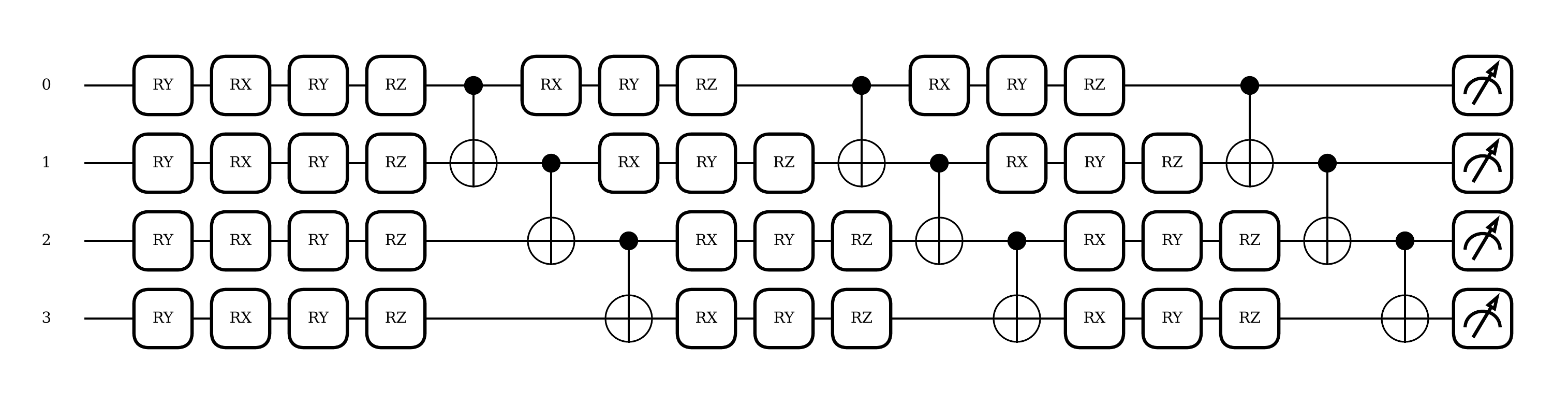}}
\caption{Illustrative hardware-efficient VQC architecture with four qubits
and three layers. In the main experiments, the number of variational
layers is specified separately for each configuration.}
\label{vqc_quantum-circuits}
\end{figure}

\paragraph*{Experimental Setup}
All experiments were conducted on the REmotely-managed
Power Aware Computing Systems and Services (REPACSS)
high-performance computing platform~\cite{repacss}. Classical
neural network components were implemented using
PyTorch~\cite{pytorch}, while quantum circuits were constructed
and simulated using PennyLane~\cite{pennylane}. Due to current
limitations in quantum hardware availability, queue times, and
the overhead associated with derivative evaluation for
physics-informed training, all training experiments were
performed in a simulator environment. Forward inference of
the trained quantum circuits may, in principle, be executed on
quantum hardware, although full physics-informed training on
hardware would require additional considerations related to
measurement noise, shot complexity, and derivative estimation.

A classical PINN baseline with four hidden layers and 32
neurons per layer was used for comparison, resulting in 6,594
trainable parameters for approximating the pressure and
stream-function fields. In contrast, for the four-qubit
configuration with ten variational layers, the FNN-TE-QPINN
and QNN-TE-QPINN architectures required 608 and 360
trainable parameters, respectively. This comparison highlights
the parameter efficiency of the quantum-assisted models.

\begin{figure}[htbp]
\centerline{\includegraphics[scale=0.22]{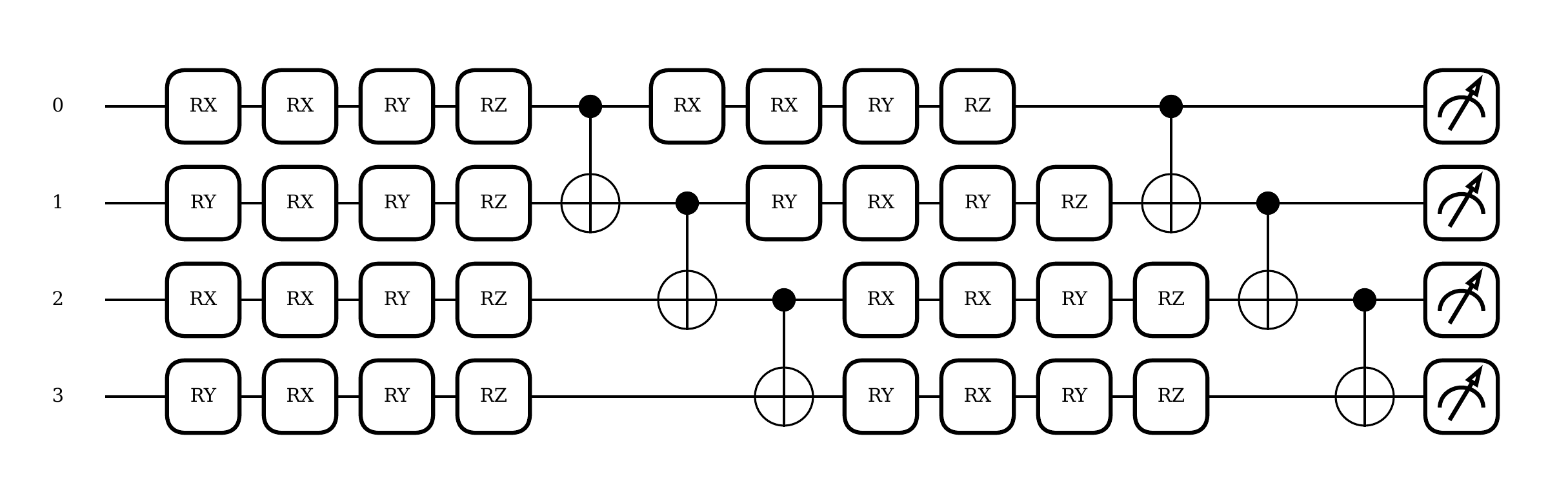}}
\caption{Illustrative QNN-based quantum embedding circuit with four qubits
and two layers.}
\label{qnn_embedding_circuits}
\end{figure}

\paragraph*{Observable and Solver Configuration}
In the experiments reported in this work, the pressure field $p(x,y)$ and
the stream-function field $\psi(x,y)$ are represented by two separate
variational quantum circuits with identical architectures but independent
trainable parameters. Each circuit employs a Pauli--$Z$-based readout
observable to produce one scalar output field. Consequently, two circuit
evaluations are required per collocation point: one for the pressure
solver and one for the stream-function solver. This design reduces
unwanted coupling between the representations of pressure and stream
function, simplifies the optimization procedure, and provides separate
control over the approximation of each physical quantity while preserving
a common architectural template across the two solvers.

\paragraph*{Quantum Circuit Architectures}
The quantum solver uses a hardware-efficient VQC composed of single-qubit rotation gates $(R_x,R_y,R_z)$ and CNOT
entangling operations, as illustrated in Fig.~\ref{vqc_quantum-circuits}.
Each VQC acts as a physics-informed quantum solver for one output field
and is applied uniformly across all collocation points.

The classical spatial coordinates are encoded into quantum states using
the QNN-based embedding circuit shown in
Fig.~\ref{qnn_embedding_circuits}. The embedding circuit maps normalized
coordinates to data-encoding rotation angles, which are then applied
before the variational evolution. The embedding circuit uses the same
number of qubits as the VQC, and its trainable parameters are optimized
jointly with the variational circuit parameters.

\begin{figure}[htbp]
\centerline{\includegraphics[scale=0.2]{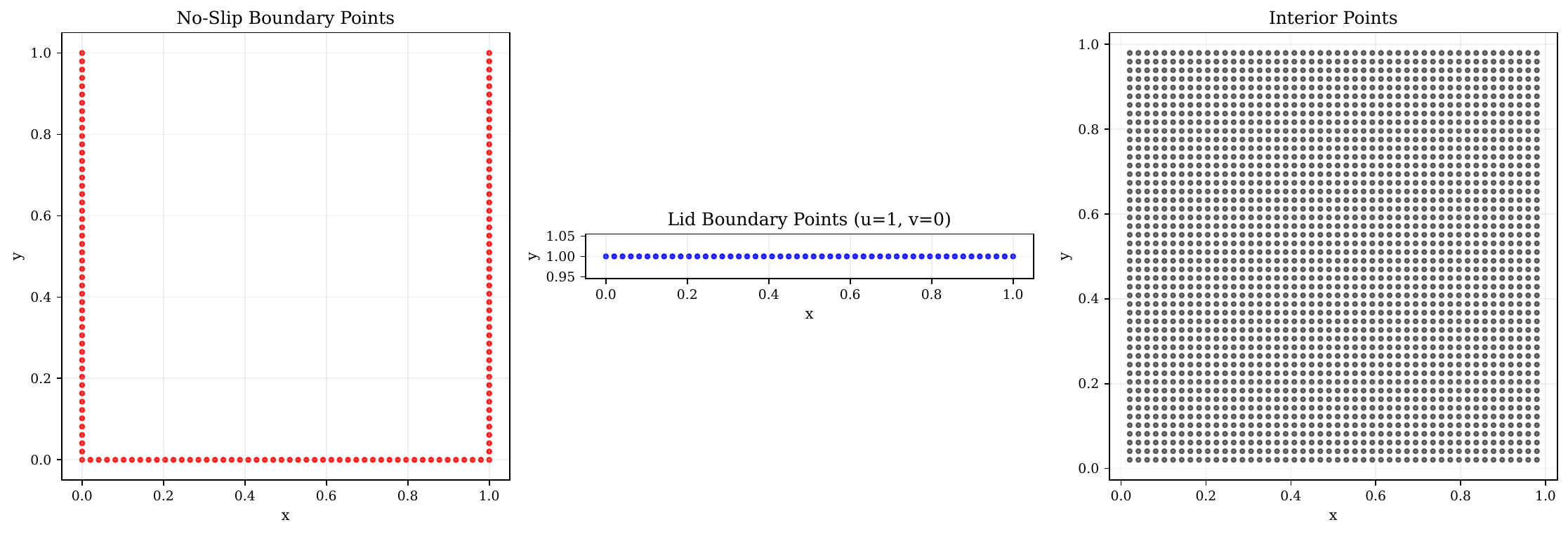}}
\caption{Distribution of collocation points for the lid-driven cavity
problem, including interior points, stationary no-slip boundaries, and
the moving-lid boundary.}
\label{collocation_points}
\end{figure}

\paragraph*{Collocation Points and Reference Solution}
The training data consist of a uniform $50 \times 50$ grid of collocation
points covering the computational domain, as illustrated in
Fig.~\ref{collocation_points}. The collocation set includes interior
points, no-slip wall boundaries on the left, right, and bottom surfaces,
and a moving-lid boundary on the top surface of the cavity. This
sampling strategy ensures that the governing equations and boundary
conditions are enforced throughout the domain in a physics-informed
manner.

\begin{figure}[htbp]
\centerline{\includegraphics[scale=0.3]{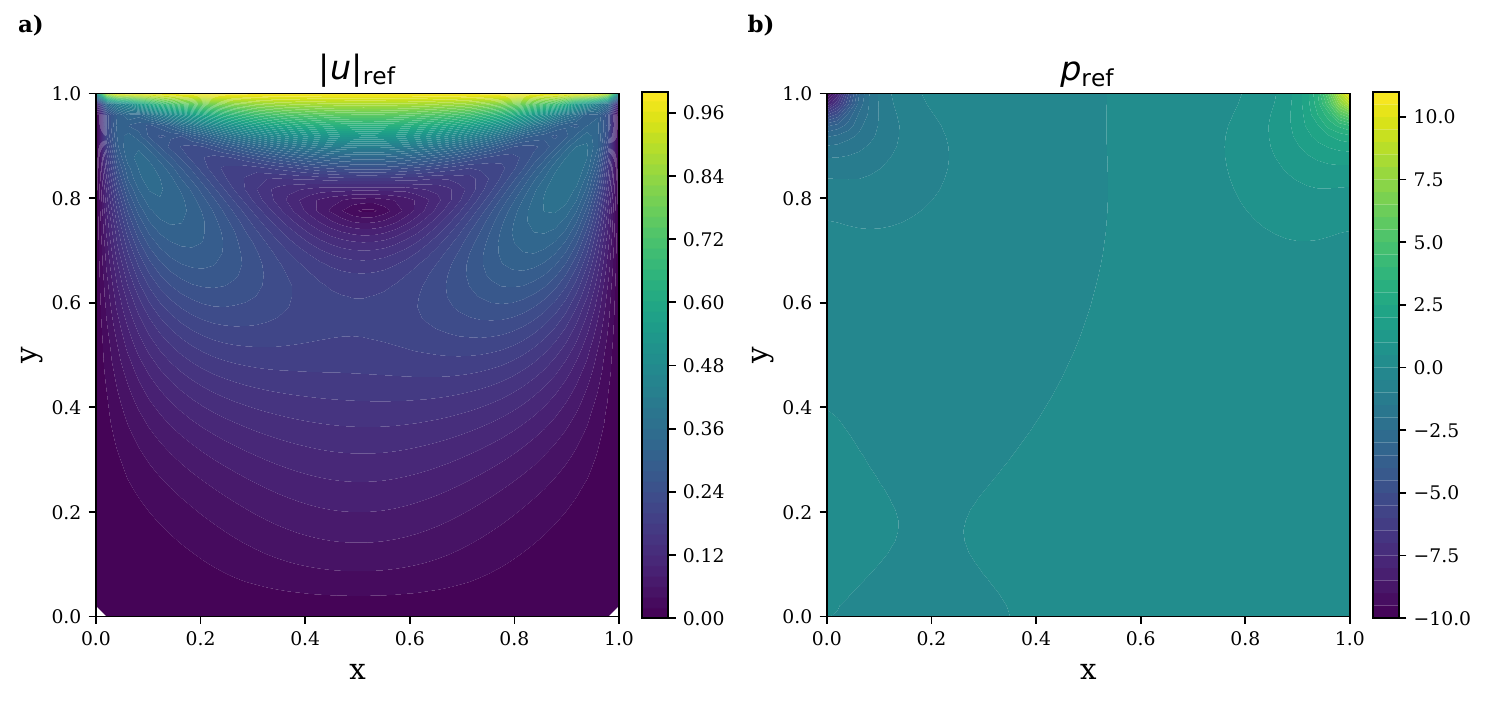}}
\caption{Reference solutions obtained using the classical RK45 solver
for (a) velocity magnitude and (b) pressure.}
\label{ref_solutions}
\end{figure}

To evaluate the accuracy of the learned solutions, reference velocity
and pressure fields were generated using a classical pseudo-time
integration approach. The steady incompressible Navier--Stokes equations
were written in pseudo-time form, spatially discretized on the
computational grid, and integrated using an adaptive Runge--Kutta
(RK45) solver until convergence to a steady state. The resulting
reference solutions for the velocity magnitude and pressure distribution
are shown in Fig.~\ref{ref_solutions} and are used as ground truth for
quantitative error analysis during training and inference.

\begin{figure}[htbp]
\centerline{\includegraphics[scale=0.3]{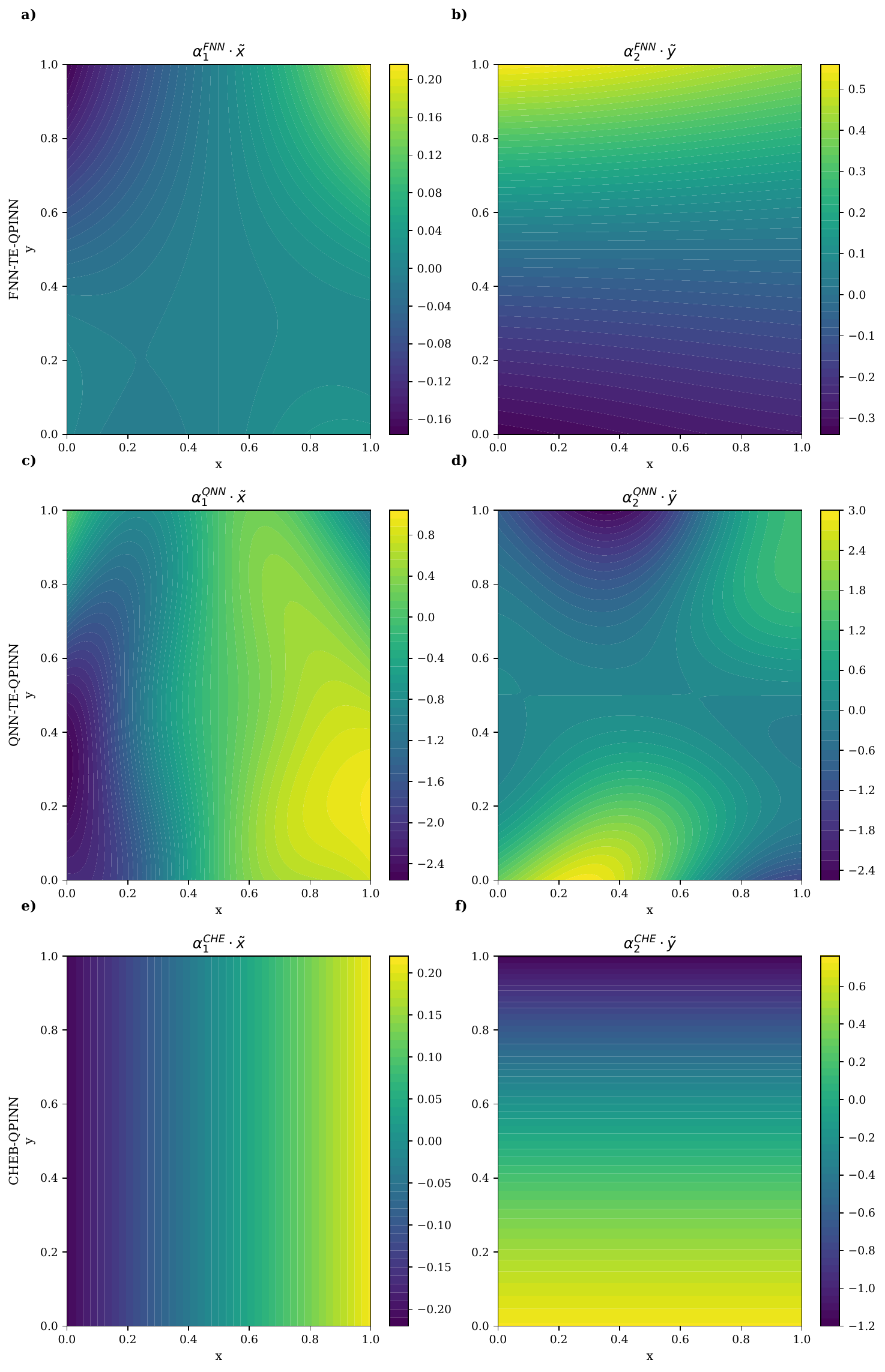}}
\caption{Coordinate-dependent quantum encoding patterns obtained from
classical collocation points for three embedding strategies:
FNN-TE-QPINN, QNN-TE-QPINN, and Chebyshev-QPINN.}
\label{embedding_results}
\end{figure}

\paragraph*{Embedding Strategy Comparison}
Since the input data for the solver consist of classical spatial
coordinates, these coordinates must be embedded into quantum states
prior to variational processing. The choice of embedding strategy plays
a critical role in determining the expressiveness and trainability of
quantum physics-informed models. To assess the impact of embedding
design, we compare three approaches: a classical feedforward neural
network (FNN)-based trainable embedding, a quantum neural network
(QNN)-based trainable embedding, and a fixed Chebyshev encoding.

Fig.~\ref{embedding_results} illustrates the coordinate-dependent
encoding patterns produced by each method. Both the FNN-TE-QPINN and
QNN-TE-QPINN embeddings yield smooth and structured representations
over the computational domain, indicating that the trainable embeddings
adapt the input feature map during optimization. In contrast, the
Chebyshev-based embedding produces more rigid patterns dominated by
predefined horizontal and vertical variations, reflecting its limited
adaptability.

These results highlight the importance of adaptive, learnable embedding
mechanisms in quantum-assisted physics-informed learning. While both
classical and quantum trainable embeddings are capable of producing
structured coordinate encodings, fixed analytical mappings such as
Chebyshev-based encodings provide less flexibility for adapting the
feature representation to nonlinear fluid dynamics problems.

\paragraph*{Training Performance}
We now examine the training performance of the proposed QNN-TE-QPINN
solver and compare it with classical PINN and FNN-TE-QPINN baselines.
All models are trained for 100 epochs using the same set of collocation
points. The Reynolds number is fixed at $Re = 10$, corresponding to a
kinematic viscosity $\nu = 0.1$, which leads to a relatively smooth
flow regime and enables stable comparison with the RK45 reference
solution.

\begin{figure}[htbp]
\centerline{\includegraphics[scale=0.23]{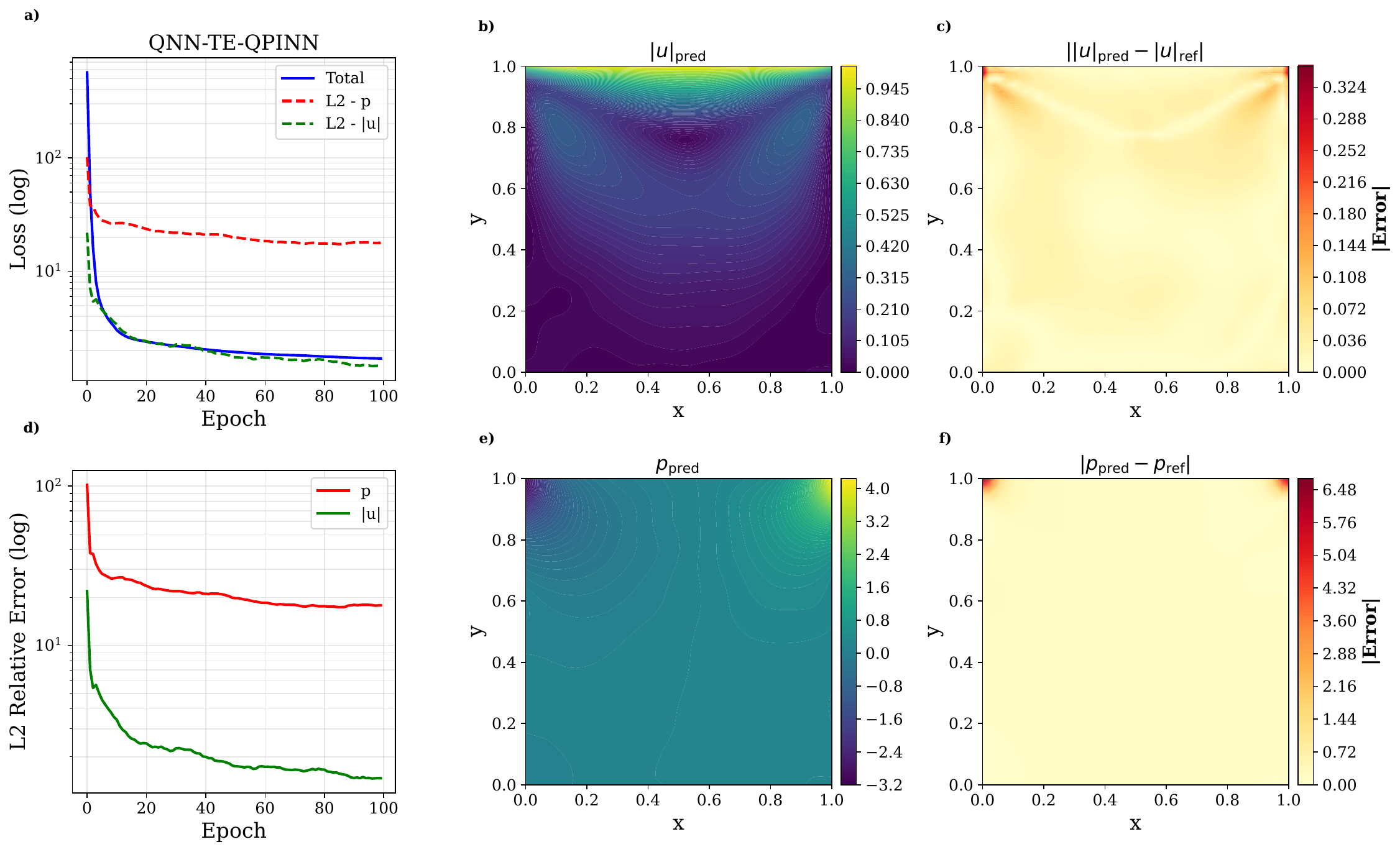}}
\caption{Training behavior of the QNN-TE-QPINN model after 100 epochs:
(a) total loss, (d) $L_2$ relative error; (b) and (e) model-predicted
velocity magnitude and pressure fields; (c) and (f) absolute error with
respect to the RK45 reference solution.}
\label{qnn_training}
\end{figure}

Fig.~\ref{qnn_training} shows the evolution of the training loss and
the $L_2$ relative error for the QNN-TE-QPINN model. Both metrics
decrease smoothly with minimal oscillations, indicating stable
optimization. The predicted velocity magnitude and pressure fields
reproduce the main features of the reference solution, while the
absolute errors remain relatively small over most of the domain,
suggesting effective enforcement of the governing equations and
boundary conditions.

\begin{figure}[htbp]
\centerline{\includegraphics[scale=0.55]{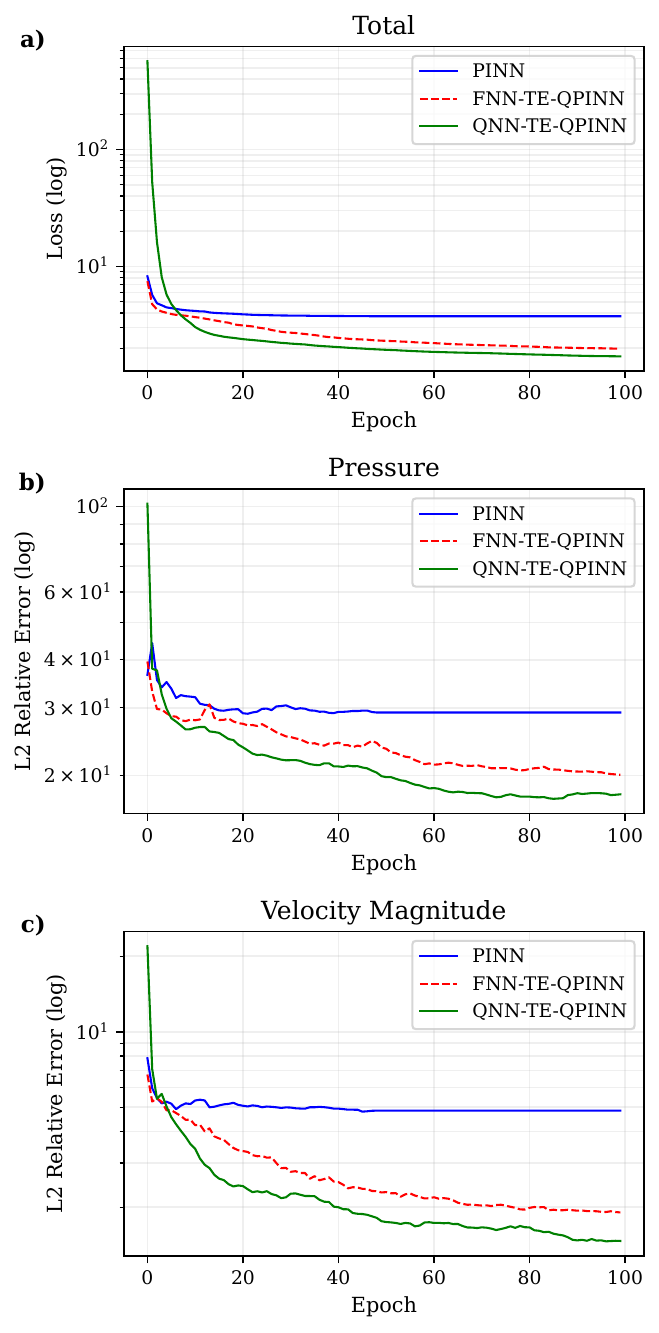}}
\caption{Training performance comparison of PINN, FNN-TE-QPINN, and
QNN-TE-QPINN: (a) total loss, (b) $L_2$ relative error of the pressure
field $p$, and (c) $L_2$ relative error of the velocity magnitude $u$.}
\label{three_comparison}
\end{figure}

Fig.~\ref{three_comparison} presents a direct comparison among PINN,
FNN-TE-QPINN, and QNN-TE-QPINN. The classical PINN exhibits early
saturation in the training loss, whereas both trainable-embedding
quantum models achieve lower loss values and improved convergence
behavior. Among the quantum-assisted models, QNN-TE-QPINN achieves
competitive performance relative to FNN-TE-QPINN while using fewer
trainable parameters.

\begin{table}[t]
\centering
\caption{Final total training loss for different embedding configurations. 
QPINN models use ten variational layers and five embedding layers, and all models are trained for 100 epochs using the same collocation set.}
\label{tab1}
\begin{tabular}{c|c|c|c}
\hline
\textbf{Qubits} & \textbf{Chebyshev-QPINN} & \textbf{FNN-TE-QPINN} & \textbf{QNN-TE-QPINN} \\
\hline
2 & 4.69 & 2.67 & 3.49 \\
4 & 2.40 & 1.99 & \textbf{1.71} \\
6 & 1.85 & \textbf{1.40} & 1.42 \\
\hline
\end{tabular}
\vspace{0mm}
\footnotesize{ \textit{*Classical PINN baseline: final loss = 2.21 under the same training setting.}}
\end{table}

Table~\ref{tab1} summarizes the effect of the number of qubits on the
final training loss of the QPINN-based models. Since the classical PINN
does not depend on the number of qubits, its values are reported only as
baseline comparisons under the corresponding training and collocation
settings. The Chebyshev-QPINN baseline uses a fixed analytical encoding
and therefore does not learn a trainable embedding map, unlike the
FNN-TE-QPINN and QNN-TE-QPINN models. The results show that the
two-qubit QNN-TE-QPINN configuration has limited representational
capacity, leading to a higher loss than the PINN and FNN-TE-QPINN
baselines, although it still outperforms the Chebyshev-QPINN
configuration. Increasing the number of qubits to four substantially
improves the QNN-TE-QPINN performance, where it achieves the lowest loss
among all reported configurations. For six qubits, both trainable-embedding
QPINN models achieve similar losses, with FNN-TE-QPINN slightly
outperforming QNN-TE-QPINN. These results suggest that the
QNN-based trainable embedding is competitive and can provide strong
parameter-efficient performance, particularly in the four-qubit
configuration.

Table~\ref{tab2} presents the final training loss under different
Reynolds numbers. The Reynolds number measures the ratio of inertial to
viscous effects,
$Re = \frac{\rho U L}{\mu} = \frac{UL}{\nu}$. At low Reynolds numbers,
viscous diffusion dominates and the flow field is relatively smooth. As
$Re$ increases, the influence of the nonlinear convective terms becomes
stronger, boundary layers become sharper, and the approximation problem
becomes more challenging for both PINN and QPINN solvers.
The results show that QNN-TE-QPINN performs competitively across the
tested Reynolds numbers. For $Re=100$ and $Re=1000$, its performance is
comparable to FNN-TE-QPINN, while for $Re \geq 3000$ it achieves the
lowest final training loss among the three models. These findings
suggest that the QNN-based trainable embedding may be beneficial in more
convection-dominated regimes, although further experiments are needed to
establish robust scaling trends.

\begin{table}[htbp]
\centering
\caption{Comparison of final total training loss for PINN, FNN-TE-QPINN,
and QNN-TE-QPINN under varying Reynolds numbers. The QPINN models use
six qubits, and all models are trained for 100 epochs using the same
collocation set.}
\label{tab2}
\begin{tabular}{|c|c|c|c|}
\hline
\textbf{Reynolds Number} & \textbf{PINN} & \textbf{FNN-TE-QPINN} & \textbf{QNN-TE-QPINN} \\
\hline
100 & 1.22 & 1.19 & 1.21 \\
1000 & 1.32 & 1.12 & 1.12 \\
3000 & 1.21 & 1.19 & \textbf{1.10} \\
5000 & 1.21 & 1.19 & \textbf{1.11} \\
10000 & 1.19 & 1.19 & \textbf{1.10} \\
\hline
\end{tabular}
\end{table}

The training results indicate that the QNN-TE-QPINN model can be trained
with relatively small quantum circuits, with the four- and six-qubit
configurations providing the most effective performance in the present
experiments. Increasing the number of qubits can improve expressive
capacity and may allow comparable performance with shallower variational
circuits, which is advantageous for NISQ devices where circuit depth is limited by noise and decoherence.
In practice, however, the experiments are constrained by the overhead of
classical quantum simulation, limiting the reported configurations to at
most six qubits or fewer than twenty variational layers, even on
high-performance computing platforms equipped with modern GPUs such as
the NVIDIA H100. The hardware-efficient circuit architecture employed in
these experiments is illustrated in Fig.~\ref{vqc_quantum-circuits}.

\begin{figure}[htbp]
\centerline{\includegraphics[scale=0.22]{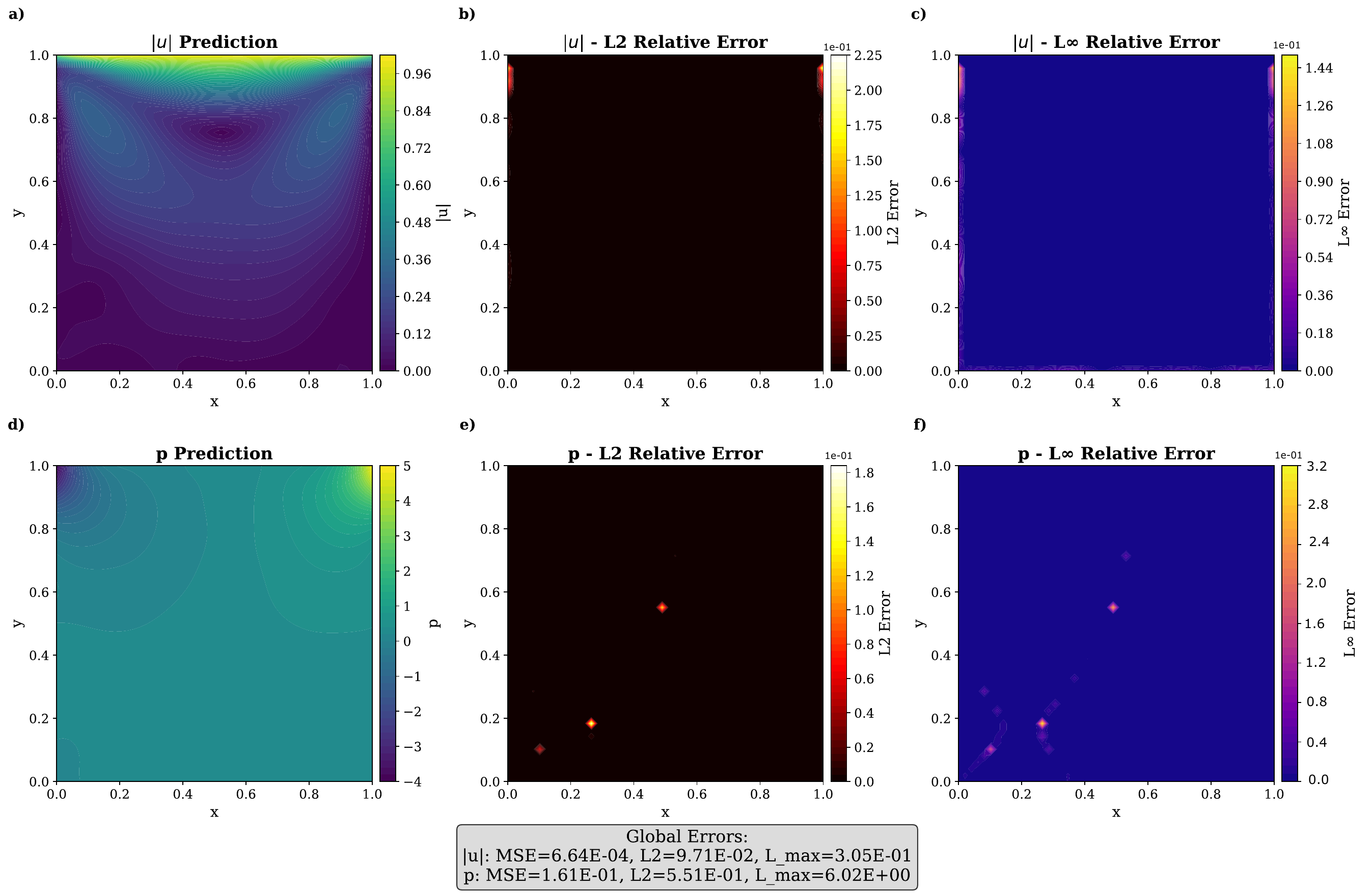}}
\caption{Inference results of the QNN-TE-QPINN solver after training for
100 epochs. (a) and (d) Predicted velocity magnitude $|\mathbf{u}|$ and
pressure $p$; (b) and (e) error distributions for $|\mathbf{u}|$ and
$p$; (c) and (f) maximum-error visualizations with respect to the RK45
reference solution.}
\label{qnn_inference}
\end{figure}

\paragraph*{Inference Results}
Following training, the generalization capability of the QNN-TE-QPINN
model was evaluated on previously unseen evaluation points. 
Fig.~\ref{qnn_inference} illustrates the inferred velocity magnitude and
pressure fields, together with their corresponding error distributions
relative to the RK45 reference solution.

The inferred velocity magnitude shows good agreement with the reference
solution, achieving a mean squared error (MSE) of
$6.64\times10^{-4}$, an $L_2$ relative error of $9.71\times10^{-2}$, and
a maximum absolute error of $3.05\times10^{-1}$. The pressure field
exhibits larger errors, with an MSE of $1.61\times10^{-1}$, an $L_2$
relative error of $5.51\times10^{-1}$, and a maximum absolute error of
$6.02$.
The larger pressure error is consistent with the difficulty of accurately
recovering pressure in incompressible Navier--Stokes solvers, where
pressure is coupled to the velocity field through the momentum equations.
In physics-informed learning frameworks, errors in spatial derivatives
of the learned stream function can propagate into the pressure residuals,
making pressure reconstruction more sensitive than velocity prediction.
Nevertheless, the predicted pressure field captures the main global
structure of the reference solution. Improving pressure accuracy remains
an important direction for future work and may require specialized loss
weighting, auxiliary constraints, or mixed formulations that treat
pressure and velocity asymmetrically.

\section{Conclusion}
\label{Conclusion}
In this work, we proposed a QNN-based trainable-embedding QPINN framework
for the steady incompressible Navier--Stokes equations in the lid-driven
cavity problem. By learning data-encoding parameters within a quantum
model, the proposed QNN-TE-QPINN provides adaptive, problem-dependent
quantum embeddings compatible with near-term quantum architectures.
Numerical results showed stable training and competitive accuracy
compared with classical PINNs and hybrid QPINNs using classical
embeddings, while requiring fewer trainable parameters. The best
performance was obtained with four qubits and ten variational layers,
where QNN-TE-QPINN improved over the classical PINN and performed
comparably to FNN-TE-QPINN. However, the fully quantum embedding did not
show a consistent advantage across all model sizes, emphasizing the need
for resource-aware architectural design.
The results indicate that QNN-based embeddings are a viable
alternative for quantum-assisted physics-informed learning. Future work
will study their scaling behavior, improve optimization strategies, and
address gradient pathologies such as barren plateaus.

\bibliographystyle{IEEEtran}
\bibliography{Main}

@misc{repacss,
  title        = {{REPACSS: High Performance Computing Center}},
  author       = {{REPACSS}},
  howpublished = {\url{https://www.repacss.org/}},
  note={Accessed in March 2026} 
}

@misc{pytorch,
  title={{PyTorch}},
  author={Paszke, Adam and Gross, Sam and Massa, Francisco and Lerer, Adam and Bradbury, James and Chanan, Gregory and Killeen, Trevor and Lin, Zeming and Gimelshein, Natalya and Antiga, Luca and Others},
  year={2019},
  howpublished={\url{http://pytorch.org}},
  note={Accessed in March 2026} 
}

@article{pennylane,
  title={{PennyLane: Automatic differentiation of hybrid quantum-classical computations}},
  author={Bergholm, Ville and Izaac, Josh and Schuld, Maria and Gogolin, Christian and others},
  journal={arXiv preprint arXiv:1811.04968},
  year={2018},
  url={https://arxiv.org/abs/1811.04968}
}

@article{ghia1982high,
  title={High-Re solutions for incompressible flow using the Navier-Stokes equations and a multigrid method},
  author={Ghia, UKNG and Ghia, Kirti N and Shin, CT},
  journal={Journal of computational physics},
  volume={48},
  number={3},
  pages={387--411},
  year={1982},
  publisher={Elsevier}
}

@article{khorasanizade2014detailed,
  title={A detailed study of lid-driven cavity flow at moderate Reynolds numbers using Incompressible SPH},
  author={Khorasanizade, Shahab and Sousa, Joao MM},
  journal={International Journal for Numerical Methods in Fluids},
  volume={76},
  number={10},
  pages={653--668},
  year={2014},
  publisher={Wiley Online Library}
}

@article{raissi2019physics,
  title={Physics-informed neural networks: A deep learning framework for solving forward and inverse problems involving nonlinear partial differential equations},
  author={Raissi, Maziar and Perdikaris, Paris and Karniadakis, George E},
  journal={Journal of Computational physics},
  volume={378},
  pages={686--707},
  year={2019},
  publisher={Elsevier}
}

@article{cerezo2021variational,
  title={Variational quantum algorithms},
  author={Cerezo, Marco and Arrasmith, Andrew and Babbush, Ryan and Benjamin, Simon C and Endo, Suguru and Fujii, Keisuke and McClean, Jarrod R and Mitarai, Kosuke and Yuan, Xiao and Cincio, Lukasz and others},
  journal={Nature Reviews Physics},
  volume={3},
  number={9},
  pages={625--644},
  year={2021},
  publisher={Nature Publishing Group UK London}
}

@article{kyriienko2021solving,
  title={Solving nonlinear differential equations with differentiable quantum circuits},
  author={Kyriienko, Oleksandr and Paine, Annie E and Elfving, Vincent E},
  journal={Physical Review A},
  volume={103},
  number={5},
  pages={052416},
  year={2021},
  publisher={APS}
}

@article{berger2025trainable,
  title={Trainable embedding quantum physics informed neural networks for solving nonlinear PDEs},
  author={Berger, Stefan and Hosters, Norbert and M{\"o}ller, Matthias},
  journal={Scientific Reports},
  volume={15},
  number={1},
  pages={18823},
  year={2025},
  publisher={Nature Publishing Group UK London}
}

@article{tran2026trainable,
  title={A Trainable-Embedding Quantum Physics-Informed Framework for Multi-Species Reaction-Diffusion Systems},
  author={Tran, Ban Q and Dehaghani, Nahid Binandeh and Aguiar, A Pedro and Wisniewski, Rafal and Mengel, Susan},
  journal={arXiv preprint arXiv:2602.09291},
  year={2026}
}

@article{tran2026quantum,
  title={Quantum-Assisted Trainable-Embedding Physics-Informed Neural Networks for Parabolic PDEs},
  author={Tran, Ban Q and Dehaghani, Nahid Binandeh and Wisniewski, Rafal and Mengel, Susan and Aguiar, A Pedro},
  journal={arXiv preprint arXiv:2602.14596},
  year={2026}
}

@inproceedings{dehaghani2025quantum,
  title={Quantum-Assisted Learning of Time-Dependent Parabolic PDEs},
  author={Dehaghani, Nahid Binandeh and Tran, Ban and Aguiar, A Pedro and Wisniewski, Rafal and Mengel, Susan},
  booktitle={2025 IEEE International Conference on Quantum Computing and Engineering (QCE)},
  volume={2},
  pages={598--599},
  year={2025},
  organization={IEEE}
}

\end{document}